\documentclass[journal=aamick,manuscript=article]{achemso}

\usepackage{chemformula} % Formula subscripts using \ch{}
\usepackage[T1]{fontenc} % Use modern font encodings
\usepackage{lineno}
\usepackage{xcolor}
\usepackage{amsmath,amssymb}
\usepackage{etoolbox}
\makeatletter
\patchcmd{\acs@contact@details}{E}{*\,E}{}{}
\makeatother

\author{Musaddaq Azeem}
\affiliation[Technical University of Liberec]
{Technical University of Liberec, Faculty of Textile Engineering, Department of Material Engineering, Studentsk\'a 1402/2, 461 17, Liberec 1, Czech Republic}
\author{Adrien Gu\'erin}
\affiliation[Universidad Adolfo Ib\'a\~nez]
{Universidad Adolfo Ib\'a\~nez, Faculty of Engineering and Sciences, Av.~Padre Hurtado 750, Vi\~na del Mar,  2562340, Chile}
\altaffiliation{Current address: Universit\'e Paris 7 Diderot, Laboratoire Mati\`ere et Syst\`emes Complexes, 75013, Paris, France}
\author{Thomas Dumais}
\affiliation[Universidad Adolfo Ib\'a\~nez]
{Universidad Adolfo Ib\'a\~nez, Faculty of Engineering and Sciences, Av.~Padre Hurtado 750, Vi\~na del Mar, 2562340, Chile}
\author{Luis Caminos}
\affiliation[Universidad Adolfo Ib\'a\~nez]
{Universidad Adolfo Ib\'a\~nez, Faculty of Engineering and Sciences, Av.~Padre Hurtado 750, Vi\~na del Mar, 2562340, Chile}
\author{Raymond E. Goldstein}
\affiliation[University of Cambridge]
{University of Cambridge, Department of Applied Mathematics and Theoretical Physics, Wilberforce Road, Cambridge, CB3 0WA, UK}
\author{Adriana I. Pesci}
\affiliation[University of Cambridge]
{University of Cambridge, Department of Applied Mathematics and Theoretical Physics, Wilberforce Road,  Cambridge, CB3 0WA, UK}
\author{Juan de Dios Rivera}
\affiliation[Pontificia Universidad Cat\'olica de Chile]
{Pontificia Universidad Cat\'olica de Chile, Departamento de Ingenier{\'\i}a Mec\'anica y Metal\'urgica, Av.~Vicu\~na Mackenna 4860, Macul 7820436, Santiago, Chile}
\author{Mar{\'\i}a Josefina Torres}
\affiliation[Pontificia Universidad Cat\'olica de Valpara\'iso]
{Pontificia Universidad Cat\'olica de Valpara\'iso, Escuela de Ingenier{\'\i}a Mec\'anica, Av.~Los Carrera, Quilpu\'e, 2430000, Chile}
\author{Jakub Wiener}
\affiliation[Technical University of Liberec]
{Technical University of Liberec, Faculty of Textile Engineering, Department of Material Engineering, Studentsk\'a 1402/2, 461 17, Liberec 1, Czech Republic}
\author{Jos\'e Luis Campos}
\affiliation[Universidad Adolfo Ib\'a\~nez]
{Universidad Adolfo Ib\'a\~nez, Faculty of Engineering and Sciences, Av.~Padre Hurtado 750, Vi\~na del Mar,  2562340, Chile}
\author{Jacques Dumais}
\email{jacques.dumais@uai.cl}
\affiliation[Universidad Adolfo Ib\'a\~nez]
{Universidad Adolfo Ib\'a\~nez, Faculty of Engineering and Sciences, Av.~Padre Hurtado 750, Vi\~na del Mar,  2562340, Chile}

\title[An \textsf{achemso} demo]
  {Optimal Design of Multi-Layer Fog Collectors}

\abbreviations{IR,NMR,UV}

\setkeys{acs}{maxauthors=20, etalmode=truncate}

\begin{document}

%\begin{tocentry}
%\includegraphics{TOCentry.pdf}
%Some journals require a graphical entry for the Table of Contents.
%This should be laid out ``print ready'' so that the sizing of the
%Inside the \texttt{tocentry} environment, the font used is Helvetica
%8\,pt, as required by \emph{Journal of the American Chemical
%The surrounding frame is 9\,cm by 3.5\,cm, which is the maximum
%permitted for  \emph{Journal of the American Chemical Society}
%graphical table of content entries. The box will not resize if the
%content is too big: instead it will overflow the edge of the box.
%
%This box and the associated title will always be printed on a
%separate page at the end of the document.

%\end{tocentry}


\newpage
\begin{abstract}
  The growing concerns over desertification have spurred research into technologies aimed at acquiring water from non-traditional sources such as dew, fog, and water vapor.  Some of the most promising developments have focused on improving designs to collect water from fog. However, the absence of a shared framework to predict, measure and compare the water collection efficiencies of new prototypes is becoming a major obstacle to progress in the field.  We address this problem by providing a general theory to design efficient fog collectors as well as a concrete experimental protocol to furnish our theory with all the necessary parameters to quantify the effective water collection efficiency.  We show in particular that multi-layer collectors are required for high fog collection efficiency and that all efficient designs are found within a narrow range of mesh porosity.  We support our conclusions with measurements on simple multi-layer harp collectors.  
  \end{abstract}


\newpage

%% INTRODUCTION
\section{Introduction}
Many regions of the world experience chronic water shortages and their associated impacts on human health and economic growth \cite{UN2018}.  This crisis has spurred research for novel technologies to exploit alternative water sources such as fog\cite{Domen2014,Schunk2018}, dew\cite{Beysens2018,Kaseke2018,Gerasopoulos2018}, and even water vapor\cite{Kim2017}. Where the conditions are favorable, fog stands out as one of the most attractive water sources because fog water can, in principle, be collected in large amounts without any input of energy \cite{Schemenauer1991,Schemenauer1994,Klemm2012}. Accordingly, a large body of work has focused on the design of efficient fog collectors \cite{Park2013,Cruzat2018,Damak2018,Holmes2015,Jiang2019,Rajaram2016,Shi2018,Zhang2015}.  Fog collection is usually achieved with fine meshes exposed to the incoming fog stream.  The minuscule fog droplets intercepted by the threads accumulate until they reach a critical size at which point the force of gravity overcomes the surface tension forces allowing the drop to slide down the collector's surface to reach the gutter at its base.  

The central design challenge for efficient fog collection involves finding the optimum balance between two physical processes that have opposite requirements \cite{Rivera2011}.  On the one hand, fog collecting meshes cannot be very dense or present a major obstacle to the flow of air otherwise the incoming fog stream will simply bypass the structure laterally.  On the other hand, fog droplets can be intercepted only if they encounter a mesh element while they transit through the collector. Therefore, meshes that are either too dense or too sparse make poor collectors.  A related issue for fog collectors is clogging of the mesh by the water droplets that have been captured thus making the collector less permeable to the incoming fog and reducing the overall water collection efficiency\cite{Park2013}.  Material scientists have sought to alleviate the problem of clogging by making structural changes to the mesh such as using harp designs \cite{Shi2018,Labbe2019} or branched patterns \cite{Andrews2011,Lin2018} instead of using the standard criss-crossing meshes that tend to trap water drops.  Other material science contributions have explored modifications of the collecting surfaces to allow intercepted droplets to coalesce and move quickly under the action of gravity\cite{Azad2015,Jing2019,Li2019}.  In particular, modifications of the contact angle hysteresis can reduce the critical size a drop needs to reach before it is freed from the mesh\cite{Park2013}.  However, many of these possible improvements will have to be scaled to realistic sizes (>1 m$^2$) and produced at a competitive price (less than \$25USD per m$^2$)\cite{Leboeuf2014} before they can be used in the field. 

\begin{figure}[H]
\includegraphics[width=0.99\textwidth]{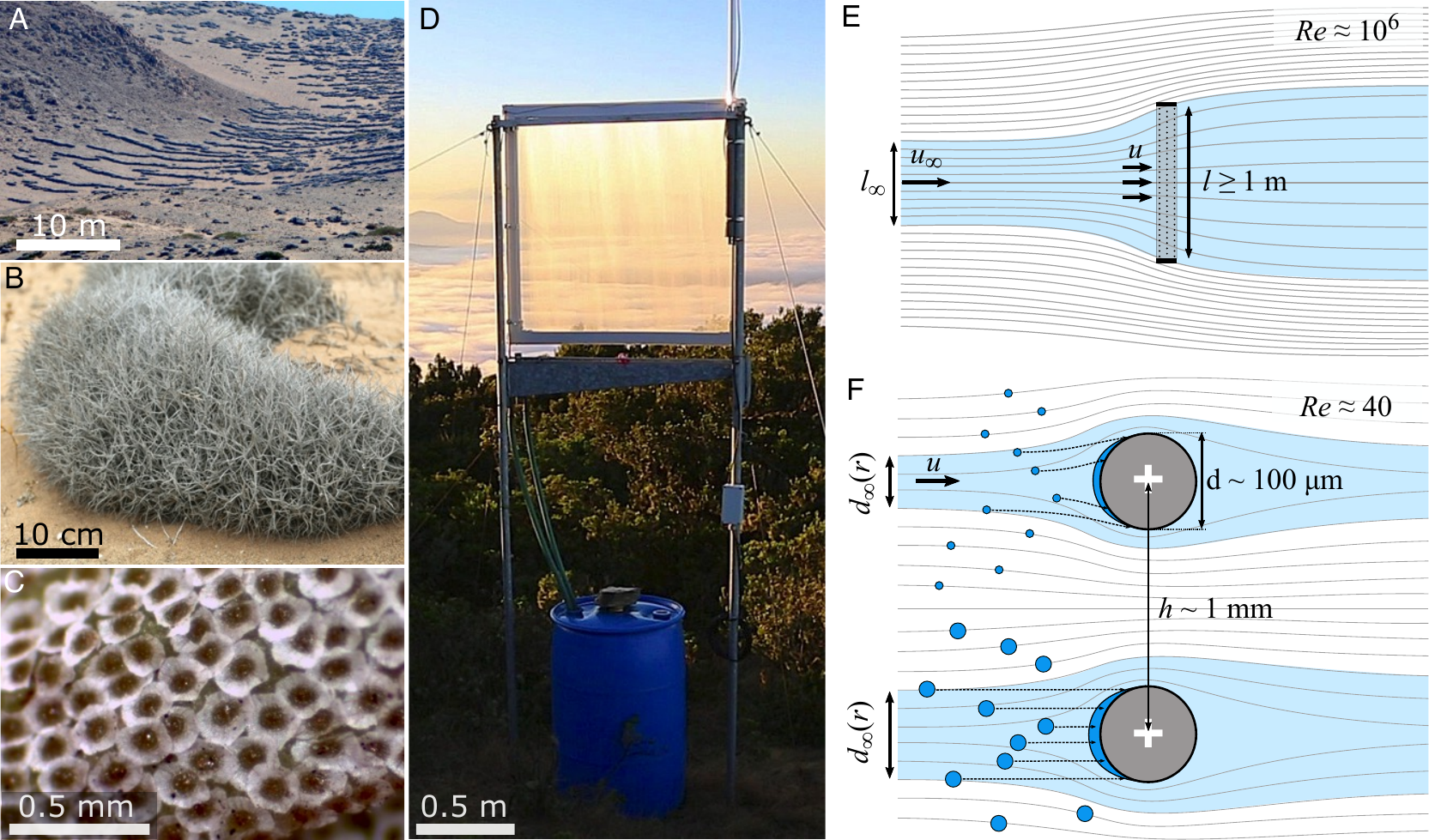}
  \caption{Aerodynamics of fog collection. (A) A stand of the bromeliad {\it Tillandsia landbeckii} in the Atacama Desert of Chile.  (B) Close-up of  {\it Tillandsia landbeckii} showing the dense three-dimensional array of leaves. (C) The hydrophilic scale-like trichomes covering the leaves and branches of {\it Tillandsia}.  (D) Prototype of a 1 m $\times$ 1 m multi-layer fog collector with a mesh solidity $s=0.3$ per layer and $N = 4$ layers.  (E) Top view of the air flow around a fog collector.  The typical collector length is  1 m $ \le l \le$ 10 m. Streamlines are drawn based on wind tunnel experiments of Ito and Garry \cite{Ito1998}, with a square mesh gauze of solidity $0.63$ at $Re =10^5$ based on the collector size.  (F) Close-up of the air flow around the section of two cylindrical threads of the collector.  The diameter of the threads $d\simeq 150-160$  $\mu$m for the collector shown in (D) and the experiments discussed below.  $d_\infty(r)$ represents the span of streamlines whose droplets of radius $r$ will be intercepted by the  thread directly downstream.  The top and bottom halves of the diagram show the interception of the small and large droplets, respectively; dashed lines indicate approximate trajectories of intercepted droplets.  Streamlines are based on Goodman's simulations \cite{Goodman1985} at $Re =20$ based on the thread diameter.}
  \label{collectors}
\end{figure}

An alternative avenue to improve the performance of fog collectors arises from observations of the bromeliad {\it Tillandsia landbeckii}, a plant that relies almost exclusively on fog to fulfill its water needs \cite{Benzing1970,Rundel1997,Raux2020}.   {\it Tillandsia} forms large stands on the fog-prone coast of the Atacama Desert of Chile.  These stands are striking in that the plants self-organize into bands orthogonal to the flow of fog (Fig.~\ref{collectors}A), thus allowing each plant direct access to the fog stream.  Moreover, the leaves and stems of {\it Tillandsia} are reduced to thin filamentous structures organized into a three-dimensional mesh, a unique feature among bromeliads (Fig.~\ref{collectors}B).  Finally, a dense layer of hydrophilic trichomes covers the plant surface (Fig.~\ref{collectors}C).  Three aerodynamically significant length scales emerge from observations of {\it Tillandsia}: the smallest length scale is that of the trichomes ($\sim100$  $\mu$m) involved in intercepting fog droplets, the intermediate length scale is the characteristic pore size between the leaves ($\sim 1$ mm) through which the fog must filter, the largest length scale is the self-organization of {\it Tillandsia} plants into fog collecting stands ($\ge 1$ m).  These observations indicate that 3-D structures, with appropriately selected length scales, can be efficient at collecting fog.  

Inspired by {\it Tillandsia landbeckii}, we investigated the potential of multi-layer fog harvesters for resolving the issues associated with single-layer collectors and improving the water collection efficiency.  Although the collectors we analyze and test do not incorporate any specific microstructure of the {\it Tillandsia} plant, their 3-D design adopts the characteristic length scales observed in these plants (Figs.~\ref{collectors}D-F). Despite having been field tested more than 50 years ago \cite{Gischler1991}, with the exception of one recent study \cite{Regalado2016}, the performance of multi-layer collectors has not been studied theoretically.  To this date, it is still unclear whether broadly applicable design principles exist.  Here, we formalize the fundamental tradeoff associated with the capture of fog with multi-layer collectors and demonstrate that simple design rules can guarantee nearly optimal fog collection efficiency.

%% FIGURE 1

\newpage

%% THEORY
\section{Theory}
\subsection*{Total water collection efficiency $\eta_{tot}$}
To formalize the performance of fog collectors, we define, as others have done before \cite{Schemenauer1989, Rivera2011, Regalado2016}, the total water collection efficiency ($\eta_{tot}$) as the water flux coming out of the collecto'’s gutter for each unit of collector area  ($J$, g$\cdot$s$^{-1}\cdot$m$^{-2}$) divided by the liquid water flux of the unperturbed fog upstream of the collector:
\begin{equation}
\eta_{tot}=\frac{J}{LWC \cdot u_\infty}~,
\end{equation}
where $LWC$ is the liquid water content of fog and $u_\infty$  is the velocity of the unperturbed fog flow, which we assume to be orthogonal to the surface of the collector.  A typical range for the $LWC$ is $0.2-0.5$ g$\cdot$m$^{-3}$ while the characteristic fog velocity  is $2-5$ m$\cdot$s$^{-1}$ \cite{Goodman1977,Schemenauer1989,Montecinos2018}.  

It is convenient to define $\eta_{tot}$ in geometrical terms by considering how a fog droplet upstream of the collector can ultimately be found in the flux of water $J$ coming out of the collector's gutter. The initial stages of collection operate at different length scales (Figs.~\ref{collectors}E,F).  First, we consider what happens at the scale of the entire fog collector (Fig.~\ref{collectors}E), where the characteristic Reynolds number based on the collector size ($l \sim$1-10 m) and unperturbed air velocity ($u_\infty \sim$ 5 m$\cdot$s$^{-1}$) is $Re = u_\infty l /\nu \sim 10^6$ ($\nu = 1.4 \times 10^{-5}$ m$^2\cdot$s$^{-1}$ is the kinematic viscosity of air).  Incoming fog droplets are part of an airstream that must filter through the collector if the droplets are to be captured.  Since the collector is an obstacle to the free flow of the airstream, a fraction of the incoming fog will simply bypass it (Fig.~\ref{collectors}E).  The filtered fraction ($\varphi$) can be quantified geometrically as the ratio of two areas: $\varphi=A_{\infty}/A$, where $A_{\infty}$  is the area of the incoming fog flow that will filter through a collector of frontal area $A$.  In the specific case of a square collector (Figs.~\ref{collectors}D,E), the filtered fraction is $\varphi = (l_{\infty}/l)^2$.
  
The second collection stage takes place at a microscopic scale and pertains to the droplets transiting through the collector.   Of these filtered droplets, only a subset will be on a trajectory that ensures collision with one of the collector elements (Fig.~\ref{collectors}F).   For any given layer of the collector, the probability that a droplet collides with a thread is given by $(d_\infty(r)/d)s$ where the ratio $d_{\infty}(r)/d$ represents the efficiency of inertial impaction for a droplet of radius $r$ (Fig.~\ref{collectors}F) and $s$ is the solid fraction, or solidity, of the layer ($s=d/h$ for our harp design).  Conversely, the probability that a droplet captured by a layer has a radius in the interval $[a,b]$ is  $s\int_a^b (d_\infty(r)/d) f(r)dr$, where $f(r)$ is the probability density function for fog droplet sizes.    Given that the mass of water provided by a droplet scales with $r^3$,  the relative contribution of droplets to the capture efficiency is $\int_a^b (d_\infty(r)/d) m(r)dr$, where
\begin{equation}
\int_a^b m(r)dr = \frac{ \int_a^b  r^3 f(r)dr}{ \int_0^\infty  r^3 f(r)dr} 
\end{equation}
is the mass fraction of liquid water contained in droplets with radii in the interval $[a,b]$ \cite{Cooper2011}. 

Finally, to these two processes, we should add the drainage efficiency ($\eta_{drain}$)\cite{Rivera2011,Regalado2016}.  The drainage efficiency represents the fraction of the intercepted volume of water that ultimately reaches the tank of the collector.  
The drainage efficiency may be reduced by re-entrainment of captured droplets under high wind conditions\cite{Gischler1991},   evaporative losses from the liquid water accumulated on the collector, and potential leaks in the gutter and pipe leading to the collector's tank.

 In the case of a single-layer collector, the three processes detailed above lead to the total water collection efficiency
\begin{equation}
\eta_{tot}=\eta_{ACE} \eta_{capt} \eta_{drain} =\underbrace{\left[ \frac{A_{\infty}s}{A}\right ]}_{\eta_{ACE}}\underbrace{\left [ \int_0^\infty \frac{d_{\infty}(r)}{d}  m(r)  dr  \right ]}_{\eta_{capt}} \eta_{drain}~,
\end{equation}  
where $\eta_{ACE}$ is the Aerodynamic Collection Efficiency (ACE) introduced by Rivera \cite{Rivera2011}. When considering a collector with $N$ layers, the total collection efficiency takes the form
\begin{equation}
\eta_{tot}= \frac{A_{\infty}}{A}  \left [ 1-\underbrace{ \int_{0}^\infty \left( 1-\frac{d_{\infty}(r)}{d}s \right)^N  m(r) dr}_{\text{lost mass fraction}}   \right ] \eta_{drain} ~,
\end{equation}  
where the term $ \left( 1-(d_{\infty}(r)/d)s \right)^N $ is the probability that a drop of radius $r$ traverses the $N$ layers  of the collector without being intercepted (see also Demoz {\it et al.} \cite{Demoz1996}).  Consequently, the integral represents the mass fraction of liquid water that filtered through the collector without being intercepted.  

Three tacit assumptions were made to arrive at Eq.~4.  These assumptions are listed here to define clearly the range of validity of our results.  First, we assume that the incoming airflow both far-field and just upstream of the collector is orthogonal to the collector's surface.  This assumption is justified because the optimum fog collectors are quite porous, with approximately 80\% of the incoming fog flow passing through the collector.  In this regime, the air velocity has a negligible component tangential to the collector surface (see below), so the interaction of the airflow with the collector filaments does not depend on position within the collector.   Second, we assume that $(d_{\infty}(r)/d)$ is constant at all locations within the collector.  This assumption implies a uniform mesh such as the harps under consideration but would have to be modified for meshes made of intersecting weft and warp threads or with threads differing in their size and shape.  Third, in deriving the lost mass fraction, we make the hypothesis that the distance between the layers is sufficiently large to allow the fog stream to regain uniformity before reaching the next layer.  As we will show below, the optimal inter-layer spacing ranges between 6 and 9 mm, which is at least 40 times greater than the characteristic thickness of the layers in our prototypes.

\subsection*{Maximizing $\eta_{tot}$}
Because Eqs.~2 and 3 are geometrical definitions of $\eta_{tot}$, they are valid irrespective of the fluid mechanics model that might be developed to quantify the collection efficiency.  Ideally, we would like to design the collector such that all steps in the harvesting of fog droplets are maximized to achieve a total water collection efficiency approaching unity.  Our goal in this section is to establish that $\eta_{ACE}$ is the only component of $\eta_{tot}$ that involves some fundamental design tradeoff. 

We begin with the drainage efficiency, $\eta_{drain}$ which is included in Eqs.~3 and 4 to take into account the possibility that captured fog droplets are either re-entrained by the airstream or otherwise lost due to leaks in the system.  Although leaks need to be taken into account in any implementation of a fog collector, they are outside the scope of our fluid mechanical analysis, but re-entrainment is not, and hence needs to be considered more carefully.  Two ways to eliminate re-entrainment are: ({\it i}) the use of multi-layer collectors to allow re-entrained drops to be re-captured by a layer farther downstream \cite{Gischler1991} and ({\it ii}) the reduction in the size of the drops clinging to the collector surface so that the drag on these drops does not exceed the critical value that would cause them to detach.  These design requirements are in fact among those put forward to optimize the other aspects of the collection process, therefore the drainage efficiency will be optimized {\it de facto}.   In what follow, we set  $\eta_{drain} = 1$ and focus on the other terms of Eqs.~3 and 4.

At the operational $Re$ number of fog collectors, the ratio $d_{\infty}(r)/d$  reflects a deposition mechanism by inertial impaction \cite{Labbe2019}.  For a droplet of radius $r$, the efficiency of impaction follows the relation\cite{Langmuir1946, Labbe2019}
\begin{equation}
\frac{d_{\infty}(r)}{d}=\frac{Stk}{Stk + \pi/2}~,
\end{equation}
where $Stk = (2 \rho_w r^2u) /(9\mu d)$ is the Stokes number, $\rho_w$ is the density of liquid water, $u$ is the velocity of the air stream, $\mu$ is the dynamic viscosity of air, and $d$ is the diameter of the thread.  This efficiency increases with increasing $Stk$; however, we note from the definition of $Stk$ that the thread diameter $d$ is the only parameter that can be tuned in the context of a passive fog collector.  Since $Stk$ increases for decreasing $d$, the width of the elements on which droplets are impacted should be reduced to a minimum.  More precisely, Labb\'e and coworkers \cite{Labbe2019} demonstrated that the size to be considered is the diameter of the thread with the water film or drops covering it.   The reduction in the size of the collecting elements can be done at constant solidity and without compromising other steps of the fog collection process.  Consequently, the geometrical ratio $d_{\infty}(r)/d$ can be made as close to 
unity as one desires, although maximizing $d_{\infty}(r)/d$ for all droplet size classes is unwarranted since the smallest droplets are the most challenging to capture and yet they represent a vanishingly small fraction of the total $LWC$ of fog \cite{Goodman1977}.

In what follows, we consider a small operating diameter for the collecting elements so that  $d_\infty \rightarrow d$.  In this limit, Eq.~4 becomes:
\begin{eqnarray}
\lim_{d_\infty \rightarrow d} \eta_{tot} = \eta_{ACE} = \underbrace{ \frac{A_{\infty}}{A} }_\varphi \underbrace{\left [ \left( 1- ( 1-s)^N \right) \right]}_\chi
\end{eqnarray}  
This equation captures in the most general form the Aerodynamic Collection Efficiency ($\eta_{ACE}$); that is, the fraction of droplets in an unperturbed upstream flow of area $A$ that are both filtered by ($\varphi$), and incident to ($\chi$), the elements of a multi-layer collector.   The ACE is of special significance because it encapsulates the fundamental trade-off in the design of efficient fog collectors.   While the incident fraction $\chi$ increases with increasing solidity $s$ and with increasing number of layers $N$, the same parameter changes reduce the collector porosity and therefore decrease the filtered fraction $\varphi$.

%% Fluid mechanical calculation of  $A_{\infty}/A$
\subsection*{Fluid mechanical calculation of  $A_{\infty}/A$}
Determining ACE for a specific collector involves finding the ratio  $\varphi = A_{\infty}/A$  using the design parameters of the collector, such as the solid fraction of the individual mesh layers and the total number of layers.  We first note that incompressibility of the flow together with mass conservation
imply $A u = A_{\infty}u_{\infty}$  (Fig.~\ref{collectors}E).  Therefore, the geometrical definition of the filtered fraction is also a statement about the ratio between the mean velocity across the collector mesh and the velocity far upstream of the collector,
\begin{equation}
\varphi= \frac{A_\infty}{A}=\frac{u}{u_\infty}~.
\end{equation}

We follow the many earlier studies of fluid flow through and around porous structures that equate two alternative definitions of the pressure drop across the porous material, the first one at the scale of the porous medium and the second one at the scale of
the far-field flow.  At the microscopic scale, the pressure drop is 
\begin{equation}
\Delta P=k \frac{\rho_{air} u^2}{2}~,
\end{equation}
where $\rho_{air}$ is the density of air and $k$ is the pressure drop coefficient for the flow of an inviscid fluid through a porous medium.   This equation arises naturally from Bernoulli's principle \cite{Cooper2011}.  As we shall see, 
since $k$ is typically not constant over a very large range of velocities,
the pressure drop coefficient is necessarily expressed in terms of the solid fraction of the medium and the Reynolds number.
At the scale of the entire collector, the pressure drop across the mesh is also related to the drag coefficient $C_D$,
\begin{equation}
    \Delta P= \frac{F_D}{A} =C_D \frac{\rho_{air} u_{\infty}^2}{2} ~,
\end{equation} 
because the drag force $F_D$ per unit area on the screen must equal the pressure drop.  Eq.~9 represents the so-called "form drag" and is valid for blunt objects at high Reynolds numbers, which is the case for fog collectors \cite{Morgan1962}.  Equating the two pressure drops, we obtain the filtered fraction
 \begin{equation}
 \varphi= \frac{A_\infty}{A}=\frac{u}{u_\infty} = \sqrt{\frac{C_D}{k} } ~.
 \end{equation}
This relation has been used in its various forms by Taylor \cite{Taylor1944,Taylor1944a}, Koo and James \cite{Koo1973}, Steiros and Hultmark \cite{Steiros2018} among many others.  

There is no consensus on how to express the drag coefficient $C_D$ and the pressure drop coefficient $k$ in terms of the design parameters of the collector mesh.  
To our knowledge, the most recent and most complete treatment is due to Steiros and Hultmark \cite{Steiros2018} (later referred to as Steiros2018); who extended the earlier work of Koo and James \cite{Koo1973} by including the "base-suction" and thus obtained accurate predictions of the drag coefficient over the entire range of solid fractions.   According to their model, the drag and pressure drop coefficients are
 \begin{eqnarray}
 C_D & = & \frac{4}{3}\frac{(1-\varphi)(2+\varphi)}{(2-\varphi)} ~, \\
 k & = &  \left(\frac{1}{(1-s)^2} - 1\right)-\frac{4}{3}\frac{(1-\varphi)^3}{\varphi^2 (2-\varphi)^2} ~.
 \end{eqnarray}

Substitution of these two relations in Eq.~10 gives an implicit relation for the filtered fraction as a function of the solidity.  Finally, because $k$ is the coefficient for the pressure drop across one layer of the collector, the total pressure drop across multiple layers is obtained by multiplying $k$ by the number of layers in the collector. The additivity of the pressure drop coefficient was shown by Eckert and Pfl{\"u}ger \cite{Eckert1942} to hold when the distance between the screens is sufficient large, and Idel'Cik showed that the pressure drop across multiple layers is additive as long as the distance of separation between the layers exceeds 15 times the size of the threads (Idel'Cik \cite{Idel'Cik1969}, page 291).

%% RESULTS AND DISCUSSION
\section{Results and discussion}
\begin{figure}[H]
\includegraphics[width=0.8\textwidth]{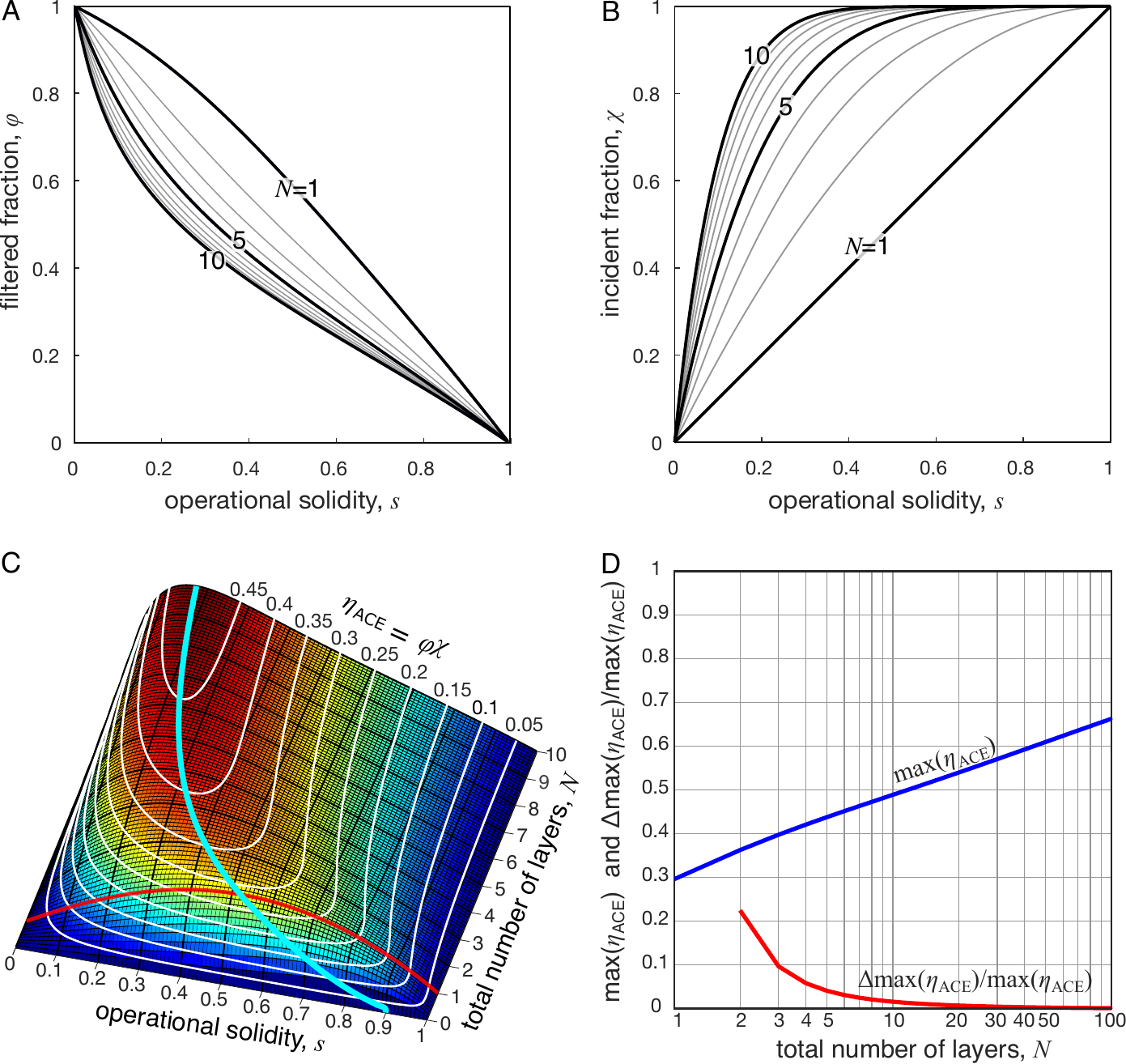}
  \caption{ Aerodynamic collection efficiency for multi-layer fog collectors.  (A) Filtered fraction predicted from the Steiros2018 model (Eqs.~10-12). (B) The incident fraction computed from geometrical considerations (Eq.~6, second term on the RHS).  (C) The ACE Ridge - a 3D representation of ACE as a function of the two control parameters $s$ and $N$.  A maximum ACE of 0.49 is observed for $10$ layers, each with an operating solidity of $0.17$.  The blue curve marks the subspace where $\eta_{ACE}$ is maximized at constant $N$.   Single-layer collectors are confined to the line $N=1$ and have an ACE below $0.3$. (Note: we have treated $N$ as a continuous variable for the purposes of illustration).   (D) The maximal ACE as a function of $N$ (plotted on a log scale).  Although max($\eta_{ACE}$) increases with increasing $N$, the relative ACE increase, $\Delta$max($\eta_{ACE}$)/max($\eta_{ACE}$), becomes small for  $N > 5$ and negligible for $N > 10$.}
\label{ACE}
\end{figure}

 To maximize the overall collection efficiency, we must seek a high filtered fraction ($\varphi$) and a high incident fraction ($\chi$).  However, these quantities are maximized at opposite ranges of the parameters $s$ and $N$ (Figs.~\ref{ACE}A,B).  The results obtained in the previous section allow us to calculate the maximum ACE found at some intermediate values of these parameters.

As can be noted in Fig.~\ref{ACE}B,  the dependence of the incident fraction $\chi$ on $N$ is highly nonlinear which gives rise to the notable advantage offered by multi-layer designs.  In a single-layer collector, the incident fraction cannot be maximized to unity, as this would imply complete obstruction of the mesh and thus
no airflow through the collector.  The use of several layers decouples, at least partially, the fluid mechanical processes behind the filtered fraction and the incident fraction.  It is therefore possible to design the collector such that nearly all upstream droplets are on a collision course with one of the collector elements while maintaining the solidity significantly below unity  (Fig.~\ref{ACE}B).  Even for a relatively modest 5-layer collector, a solidity as low as 0.5 can already guarantee a near maximal incident fraction  (Fig.~\ref{ACE}B).  The possibility of greatly increasing the incident fraction for intermediate solidity values is the reason why multi-layer collectors can be made much more efficient. Moreover, since the equation for the incident fraction is purely geometrical, there is no doubt about the general validity of this conclusion. 

Computation of the aerodynamic collection efficiency  $\eta_{ACE}=\varphi \chi$ for a broad parameter range  indicates that it reaches a maximum of  49\% for $N=10$ (Fig.~\ref{ACE}C).  In contrast, single-layer collectors are confined to the line $N=1$ and can reach a maximal ACE of only 30\% at an operational solidity slightly above 0.5. Increasing the number of layers beyond 10 increases the ACE further; with the theoretical possibility of reaching an ACE of unity for very large $N$ (Fig.~\ref{ACE}D).  This limiting behavior raises the question of how many layers should be used  in practice.   An answer emerges when considering the contribution to the total ACE made by each new layer (Fig.~\ref{ACE}D).   Beyond $N=5$, the relative increase in ACE becomes vanishingly small. Therefore, considerations about the most efficient use of available materials would suggest that the number of layers should be limited to approximately 5, at least in the limit where $d_\infty \rightarrow d$.

%% FIGURE 2

\begin{figure}[h]
\includegraphics[width=1\textwidth]{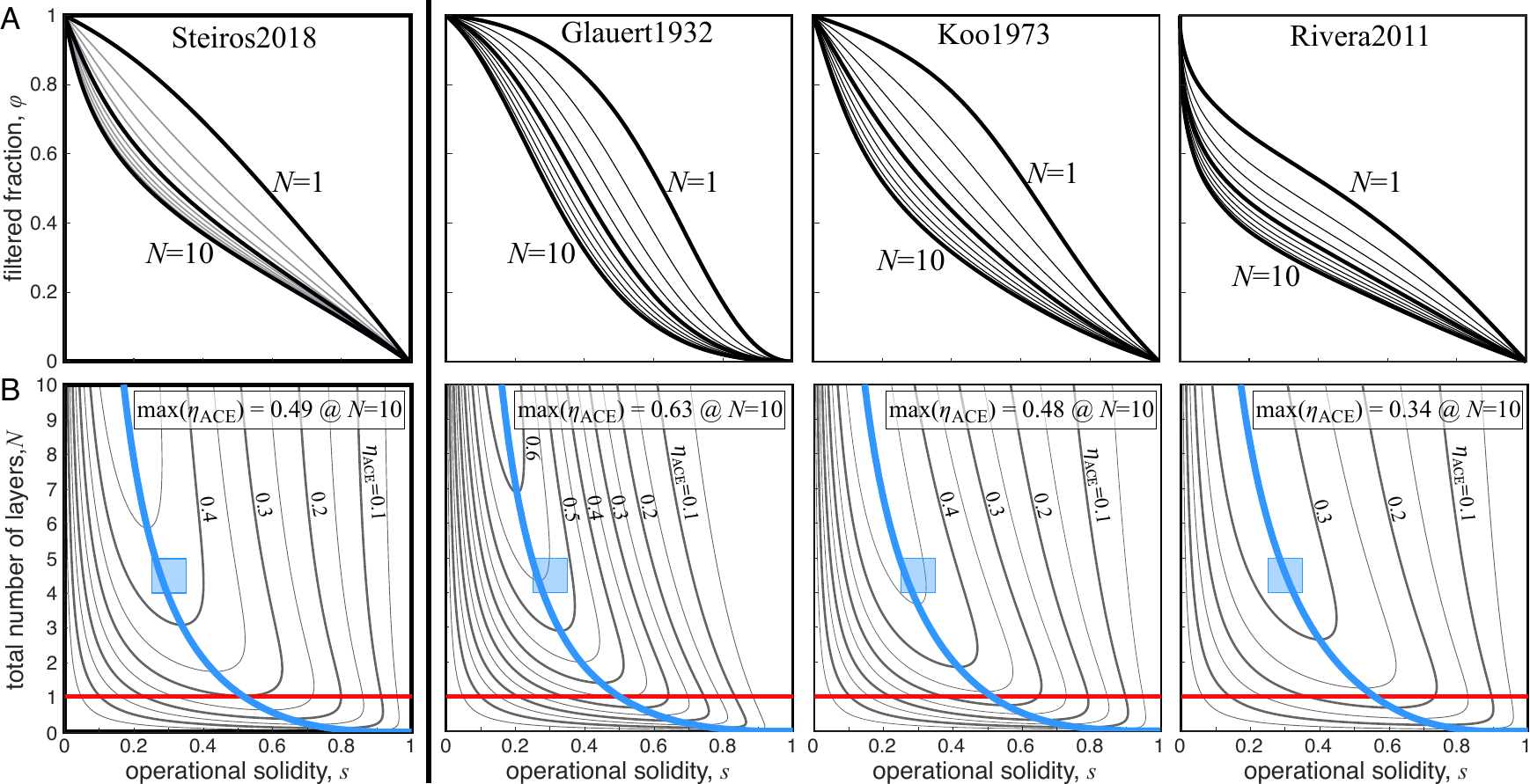}
\caption{ Comparative analysis of the ACE ridge.  (A) The filtered fraction predicted by four fluid mechanics models. Note the model-dependent form of the asymptotic behavior of $\varphi(s)$ as $s\to 0$. (B) Design space for the models listed in (A). The blue curve marks the subspace within which ACE is locally maximized at constant $N$. The blue square is the suggested target design. The red line at $N=1$ is the design space for single-layer collectors.}
\label{models}
\end{figure}

As indicated in the theory section, the Steiros2018 model is one of many 
models \textemdash published over a period of 80 years\textemdash ~that provide a fluid mechanical formulation for the filtered fraction (Suppl.~Mat).  The functional form as well as the asymptotic behavior of the filtered fraction predicted by alternative theories vary substantially (Fig.~\ref{models}A).  In that respect, the Glauert1932 model \cite{Glauert1932} and the Rivera2011 model \cite{Rivera2011} represent two extreme behaviors, while the Steiros2018 model \cite{Steiros2018} adopted here and its precursor, the Koo1973 model \cite{Koo1973}, are intermediate for the limiting behavior of $\varphi$ as $s \rightarrow 0$.  The prediction of the models for small solidity is especially important in the context of multi-layer collectors since their maximal ACE is attained for solid fractions below 0.3 (Fig.~\ref{models}B).   

A comparative analysis of the design space for these models is also informative.  Notably, although the models disagree on the maximum $\eta_{ACE}$ that can be achieved for a given $N$, their respective ACE ridges follow similar arcs in design space 
(Fig.~\ref{models}B).  Specifically, they all go through a small target area ($0.25<s<0.35$, $N=4,5$) where the multi-layer collectors achieve an efficiency $\sim\! 40$\% better than the most efficient single-layer collectors.   The quantitative agreement between the models shows the robustness of the efficiency optimization in design space (see also Regalado and Ritter \cite{Regalado2016} for qualitatively similar results).  Interestingly, the subspace where $\eta_{ACE}$ is locally maximized follows closely curves of constant filtered fraction for all four models (Fig.~S1).  Therefore, the improved aerodynamic collection efficiency of multi-layer fog collectors comes almost exclusively from improvements in the incident fraction as new layers are added to the system.  

%% FIGURE 3

\begin{figure}[h]
\includegraphics[width=0.8\textwidth]{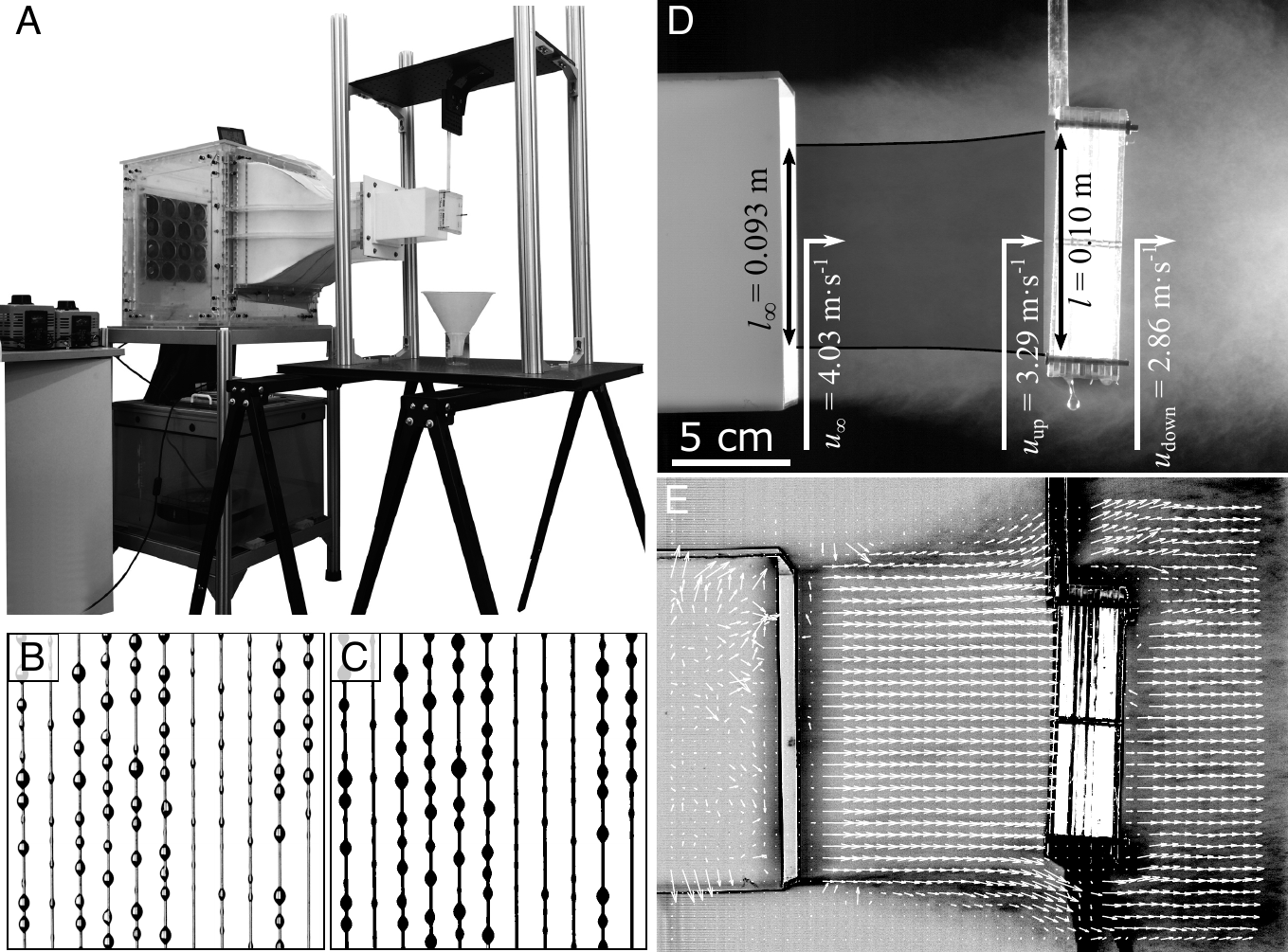}
  \caption{Measurement of ACE for a multi-layer harp collector ($s = 0.17$, $N = 4$).  (A) Fog tunnel with $14$ cm $\times$ $14$ cm working section. (B) Photo of the mesh under operating conditions ($h=2$ mm, $d=0.150$ mm).  (C) Binary (black/white) version of (B) used to compute the solidity.   The "dry" solidity is $0.075$ while the "wet", operational solidity is $0.17$.   (D) Close-up of the fog jet filtering through the collector with the key variables characterizing the flow field indicated. (E) Detailed flow field used to infer the variables in (D). (see Suppl.~Mat.~for movie)}
 \label{measuredACE}
\end{figure}

Because the models differ substantially in their predicted maximum ACE (from 34\% to 63\% for a 10-layer collector), we undertook a series of experimental observations to quantify the efficiency of multi-layer collectors.   As noted above, the equation for $\eta_{ACE}$  is, first and foremost, a statement about two geometrical ratios: the area ratio associated with the filtered fraction and the solidity $s$  of the mesh (ratio of obstructed area over the total area of one collector layer).  To assess the ACE, we developed a wind tunnel to produce realistic fog conditions in the laboratory (Fig.~\ref{measuredACE}A, Suppl.~video).  Experimenting with a  4-layer harp collector ($l=100$ mm,  $h=2$ mm, $d=0.150$ mm), we found an operating solidity of $s=0.17$ (Figs.~\ref{measuredACE}B,C), giving an incident fraction of $\chi=1-(1-s)^4 = 0.53$.  Integrating the flow field, we arrived at a filtered fraction of $\varphi_{obs} = (l_\infty/l)^2= 0.81\pm0.016$ (Figs.~\ref{measuredACE}D,E).   Based on the measured incident and filtered fractions, the aerodynamics collection efficiency is  $\eta_{ACE}=\varphi\chi = 43\%$, which exceeds slightly the value of 37\% predicted by the Steiros2018 model (Fig.~\ref{ACE}C).  The discrepancy arises in part because of the impossibility of measuring the flow field within 10 mm of the collector's surface with our current experimental set-up. The truncated velocity field leads to a slight overestimate of the filtered fraction (Table S1, Fig.~S2).  A more complete reconstruction of the velocity field could be achieved with other flow visualization methods such as the smoke-wire technique\cite{Batill1981}.

%% FIGURE 4

Given the care needed to measure ACE, it might be asked why it should be preferred as a performance standard over the total water collection efficiency $\eta_{tot}$ as defined in Eq.~1. The reason is that although Eq.~1 appears tractable at first sight, a more detailed analysis (Eq.~4) reveals that $\eta_{tot}$ involves the lost mass fraction,
$ \int_{0}^\infty \left( 1-(d_{\infty}(r)/d)s \right)^N  m(r) dr~, $
where the terms $d_{\infty}(r)/d$ and $m(r)$ both depend on the radius of the droplets in the incoming fog.   Notably, these two terms give, together, a scaling on the order of  $r^5$ (see the Theory section).   Therefore, unless the probability density function for the droplet sizes, $f(r)$, is characterized precisely, the total water collection efficiencies are impossible to compare.  In fact, it could be argued that due to its very nonlinear dependence on $r$, $\eta_{tot}$ is not appropriate as a metric for efficiency because of its great sensitivity to the presence of rare but large droplets.  
In contrast, ACE is what is left of  $\eta_{tot}$ when factors affected by the droplet size distribution of fog are eliminated (Eq.~6), and, moreover, it captures the fundamental trade-off for fog collection.  Therefore, in an effort to increase the repeatability and portability of future research in fog collection, we propose the geometrical measurement of ACE as a potential standard for the field (Fig.~S3).

\begin{figure}[h]
\includegraphics[width=1\textwidth]{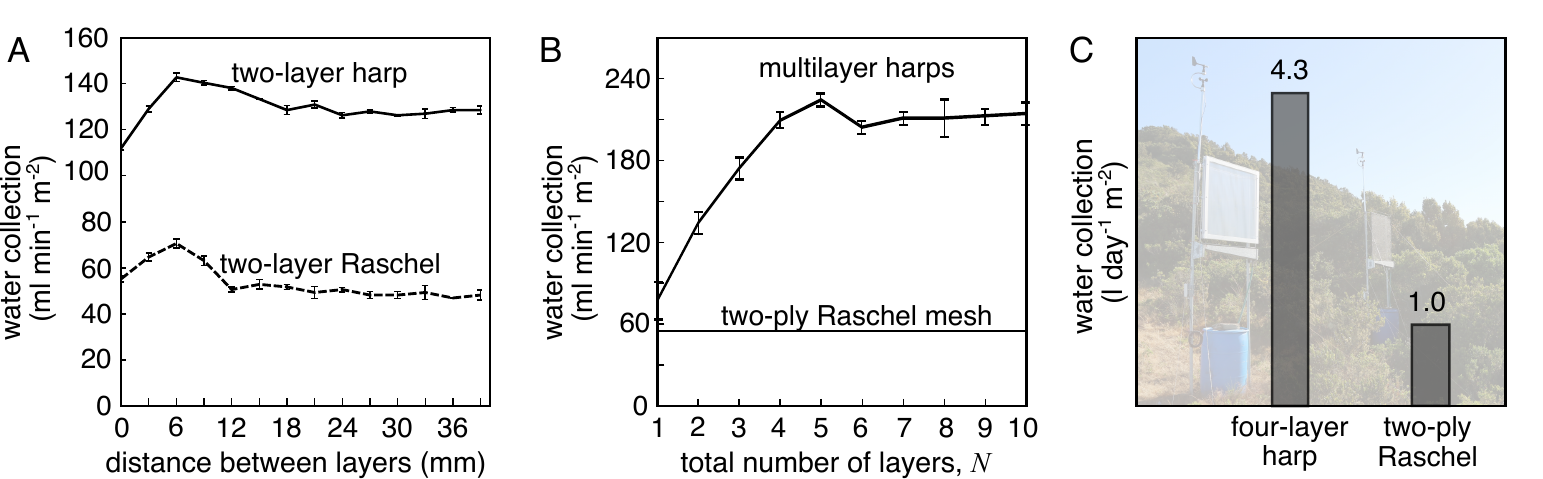}
\caption{Yield measurements.  (A) Effect of inter-layer spacing on the yield of multi-layer collectors.  (B) Yield of multi-layer harps ($1\le N \le 10$, $s=0.17$, inter-layer spacing of 6 mm) compared to two plies of Raschel mesh with $s=0.7$ at a fog velocity $u_\infty=4$ m$\cdot$s$^{-1}$. (C) Field measurements of yield over 20 days.}
 \label{yields}
\end{figure}

As a final validation of the performance of multi-layer collectors, we compare their yield with that of the standard fog collecting medium - two plies of Raschel mesh ("dry" solidity $s=0.6$) \cite{Schemenauer1994.2} without spacing between them and thus approximating a single-layer collector.  As expected, the yield of the multi-layer harps greatly exceeds that of the Raschel standard (Fig.~\ref{yields}).  Notably, even a single harp layer offers a slightly better yield than the two-ply Raschel mesh (Fig.~\ref{yields}B).   The poor performance of the Raschel mesh under well-defined laboratory conditions is explained by the fact that the two-ply mesh exceeds greatly the optimal operational solidity ($s_{Raschel} \simeq 0.7$ vs $s_{opt} \simeq 0.5$).  While the multi-harp designs outperform single-layer designs for all $N$, these collectors lose some of their yield for $N \ge 6$ (Fig.~\ref{yields}B), a result that is not predicted from the design space.  This efficiency loss probably arises because of the increasing boundary layer that develops in the vicinity of the collector frame.  In the case of a 10-layer collector, the frame depth exceeds 50 mm while the open area for filtration remains 100 mm $\times$ 100 mm.  In other words, for large $N$, the collector depth is such that the collector forms an increasingly long tube through which the fog stream must flow.  Despite this limitation, the five-layer harp offers a four-fold increase in yield (Fig.~\ref{yields}B).  These results were confirmed in field experiments with the 4-layer harp prototype shown in Fig.~\ref{collectors}D.  During a period of low fog, the prototype collected 4.3 l$\cdot$day$^{-1}\cdot$m$^{-2}$ while the two-ply Raschel mesh collected only 1 l$\cdot$day$^{-1}\cdot$m$^{-2}$ (Fig.~\ref{yields}C).

%% FIGURE 5

\section{Conclusions}
In this paper, we have presented designs for optimally efficient passive fog collectors by focusing on a 
geometrical relation (Eq.~6) known as the aerodynamic collection efficiency (ACE).  As we have shown, the maximal values of ACE are achieved only through the use of multi-layer collectors whose efficiency can exceed by 40\% that of the best single-layer collectors. The analysis shows that, taking into account the most effective use of materials, the optimal fog collector has $N = 4,5$ layers and operating solidity $s=0.3\pm0.05$, assuming that the {\it operating} thread diameter is sufficiently small to maximize inertial impaction of fog droplets.  These conclusions were validated experimentally for multi-layer harp collectors.  When optimized,  the latter can collect as much as four times that collected by the standard two-ply Raschel mesh, both under laboratory and field conditions.

%% EXPERIMENTAL
\section{Experimental}
{\bf Collector design - }Multi-layer collectors were built using fast prototyping tools.  Using a laser cutter (Ready Cut), square plexiglass frames with a 100 mm $\times$ 100 mm central open area were fabricated.  Evenly spaced notches (typical spacing: 1 mm $\le h \le$ 2 mm) were made in the upper and lower edges of the frame to hold polyethylene monofilaments ($d$ = 150-160 $\mu$m) into a vertical harp arrangement.  These frames were then stacked with different inter-layer spacings to form multi-layer fog collectors. The experiments reported here were done with a staggered relative alignment between successive layers.   Note, however, that the staggered or in-line arrangements of layers had no significant effect on the performance of the collector.

\noindent{\bf Yield measurements -} To measure the yield, the prototypes were hung at a distance of 100 mm from the opening of a wind tunnel equipped with a fog chamber (see below).  The water intercepted by the mesh was collected in a funnel leading to a graduated cylinder.   Collection occurred over a total time interval of $15$ min following an initial saturation period of $5$ min.

\noindent{\bf Measurement of the aerodynamic collection efficiency -} Flow experiments were performed with an open-jet wind tunnel developed specifically to measure the efficiency of fog collector prototypes under natural conditions.  The tunnel consists of two elements: a lower nebulization chamber for fog production and an upper flow chamber to accelerate the fog cloud and guide it into a uniform jet (Fig.~\ref{measuredACE}A).  The nebulization chamber contained $\sim 50$ liters of water within which was immersed a 300 W $12$-head ultrasonic nebulizer (Model DK12-36).  The fog produced in this chamber was injected into the upper chamber using a $16$ W, 200 mm $\times$ 200 mm ventilation fan.  Within the flow chamber, an array of $16$, 80 mm $\times$ 80 mm, computer fans accelerated the fog towards a contraction that converged the fog stream to a jet of 140 mm $\times$ 140 mm in cross-section.  Both the ventilation fan and the array of computer fans were powered through variable voltage transformers allowing us to set the jet velocity in the range $0.1-4.2$ m$\cdot$s$^{-1}$.  A honeycomb filter was placed at the upstream end of the contraction to eliminate turbulence and provide a homogeneous fog flow.

The flow of fog through and around the collector prototypes was visualized using a Phantom V611 high speed camera equipped with a Canon EF $100-400$mm telephoto zoom.  Images were acquired at a rate of $4000$ fps (exp.~240 $\mu$s) with a camera resolution of $1024\times 768$ pixels and an image scale of 270$\mu$m/pixel.  Analysis of the flow pattern was performed using a Matlab program first developed by Dr.~A.F.~Forughi at the University of British Columbia (Vancouver, Canada) and made freely available on Github (https://github.com/forughi/PIV).

\begin{acknowledgement}
MA thanks the Technical University of Liberec (TUL) for a Student Grant (SGS 21313) 2019.  JD acknowledges funding from Fondef (ID15i10387) and Fondecyt (1130129).  REG and AIP thank the Engineering and Physical Sciences Research Council (UK) for support under grant EP/M017982/.
\end{acknowledgement}

\begin{suppinfo}
The following files are available free of charge.
\begin{itemize}
  \item Supplementary Material: Table S1, Figures S1-S3, description of alternative fluid mechanics models for the filtered fraction.
  \item Harp movie: movie of the fog flow through a 4-layer harp collector.
\end{itemize}

\end{suppinfo}


\bibliography{Collectors}

\makeatletter
\patchcmd{\acs@contact@details}{E}{*\,E}{}{}
\makeatother

\newcommand{\beginsupplement}{%
        \setcounter{table}{0}
        \renewcommand{\thetable}{S\arabic{table}}%
        \setcounter{figure}{0}
        \renewcommand{\thefigure}{S\arabic{figure}}%
     }
     
     \newpage
     
     \section{Supplementary Material}
\beginsupplement

% FIGURES
\section*{Tables and Figures}

%% TABLE 1

\begin{table}
\centering
%\caption{
{Table S1: The filtered fraction $\varphi$ computed as a ratio of areas  ($l_\infty^2/l^2$).}
%\begin{tabular*}{\hsize}{@{\extracolsep{\fill}}lccc} 
\begin{tabular}{@{\vrule height 10.5pt depth4pt  width0pt}lcccccc}
\hline
Collector     & $l_\infty$  &  $l$   & $l_\infty^2 / l^2 $  \\
\hline
four-layer harp  &      0.093    &   0.10    &     0.82   \\
closed           &	   0.047    &   0.10     &     0.21   \\
open            &	  0.096     &   0.10     &    0.88 \\
\hline 
\end{tabular}
\label{filteredfraction}
\end{table}

%\begin{table}
%\centering
%\caption{The filtered fraction, $\varphi$, computed as a ratio of velocities ($u/u_\infty$) or areas  ($l_\infty^2/l^2$).}
%%\begin{tabular*}{\hsize}{@{\extracolsep{\fill}}lccc} 
%\begin{tabular}{@{\vrule height 10.5pt depth4pt  width0pt}lcccccc}
%\hline
%Collector   &  $u_\infty$  &  $u$  & $u/u_\infty$  & $l_\infty$  &  $l$   & $l_\infty^2 / l^2 $  \\
%\hline
%4-layer harp  &  4.03	  &	3.29	 &	0.82     &      0.088    &   0.098    &     0.82   \\
%closed          & 3.99	 &	1.33	 &	0.33	   &	   0.043    &   0.098     &     0.20   \\
%open            & 4.04	 &	3.63	 &	0.90	   &	  0.094     &   0.098     &    0.92 \\
%\hline 
%\end{tabular}
%\label{filteredfraction}
%\end{table}

\newpage

%% FIGURE S1
\begin{figure}[H]
\includegraphics[width=1\textwidth]{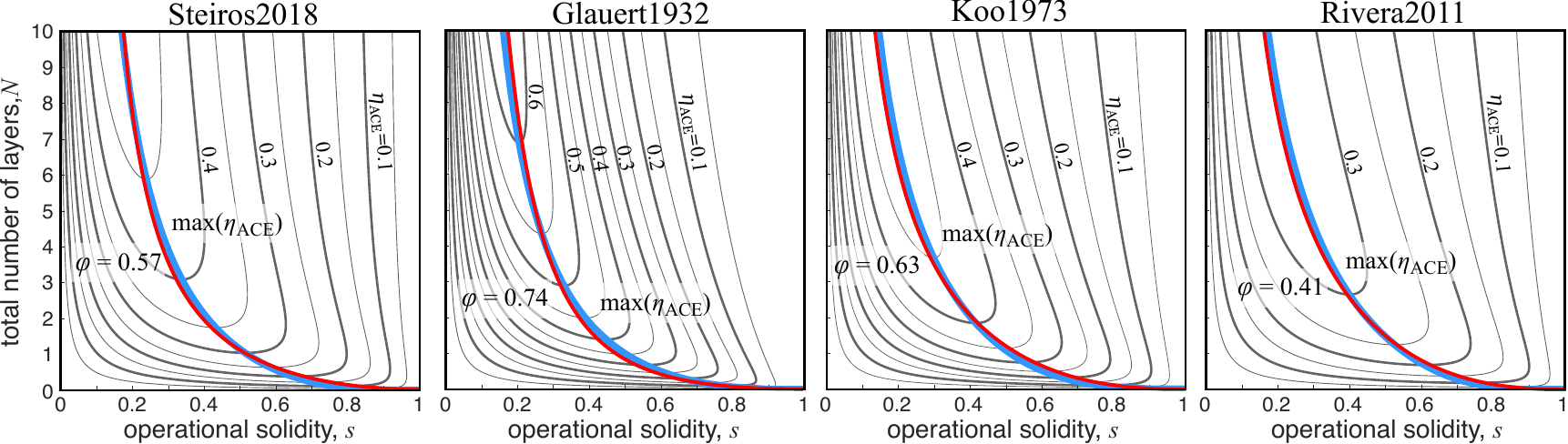}\\
\caption{The max($\eta_{ACE}$) subspace (blue curves) overlaps closely with level curves for the filtered fraction  (red) in design space.}
 \label{S1}
\end{figure}

\newpage
%% FIGURE S2
\begin{figure}[H]
\includegraphics[width=0.4\textwidth]{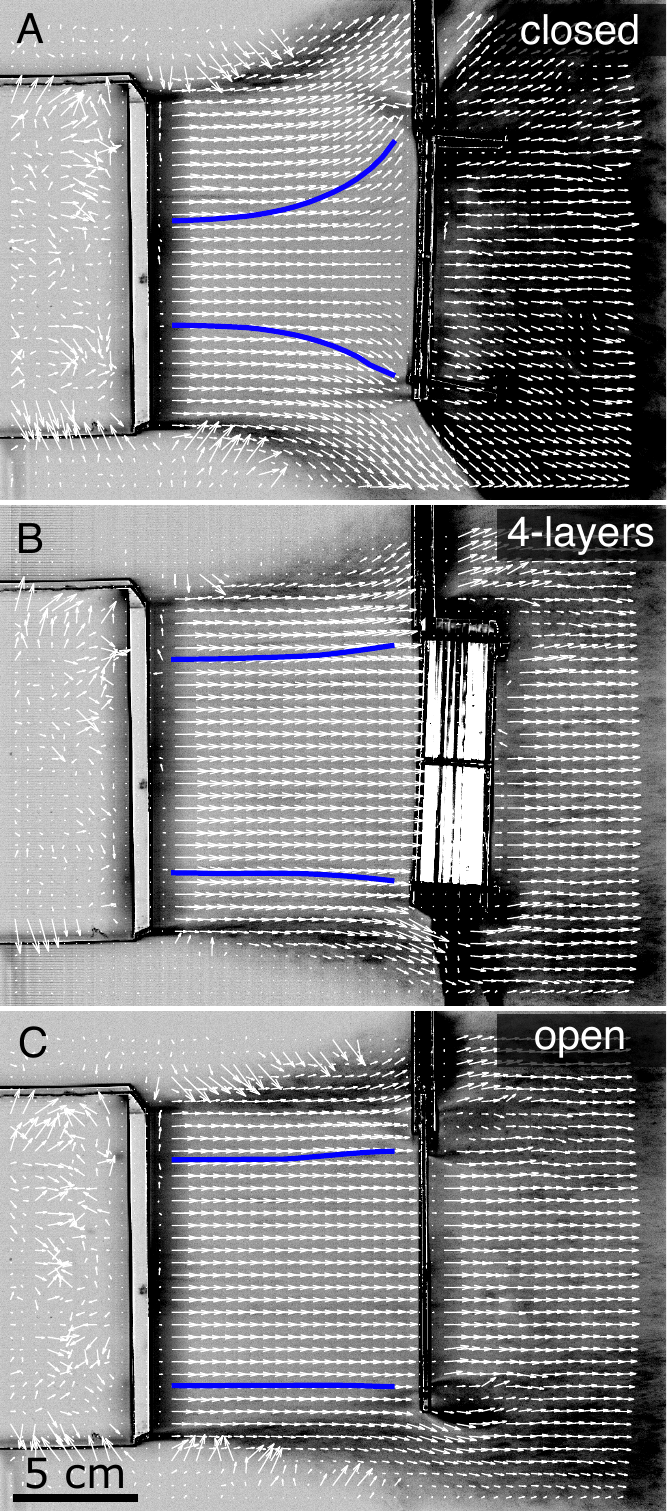}\\
\caption{Fog flow for three test conditions: a closed collector (A), a four-layer harp collector (B), and an open collector where only the frame obstructs the flow (C).  In all three cases, the  ``collector'' was square with a central area of 100 mm $\times$ 100 mm and a frame of 7 mm on all four sides. The blue curves show the streamline dividing the filtered flow from the by-pass flow.   The flow field downstream of the closed collector is not zero because the visualization protocol captures the flow that has by-passed the solid surface laterally. Also, the area ratio in (A) is not zero because our protocol to map the flow field does not capture the flow within 10 mm of the collector surface.  This effect leads to an artificially inflated filtered fraction.}
 \label{S2}
\end{figure}

\newpage
%% FIGURE S3
\begin{figure}[h]
\includegraphics[width=0.8\textwidth]{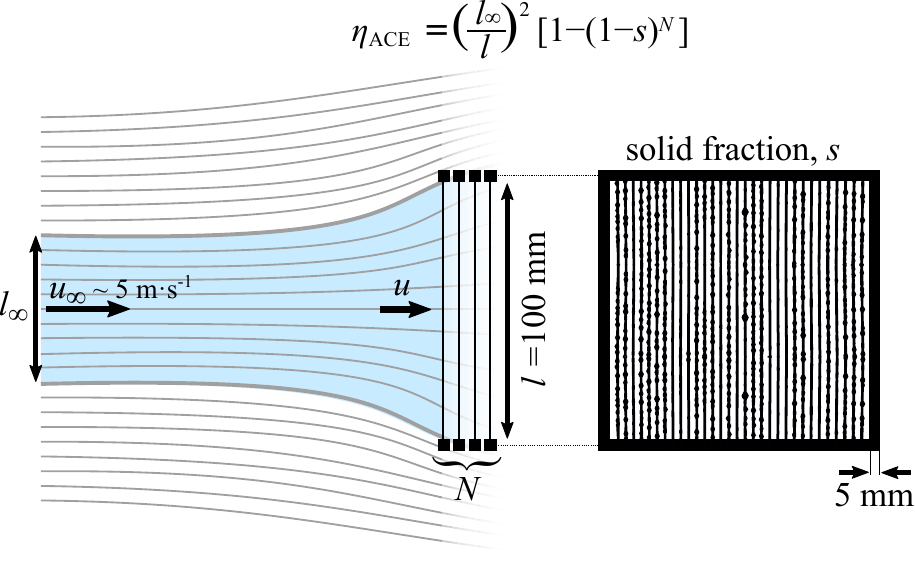}\\
\caption{Proposed standard for the measurement of ACE. Prototypes should be square with 100 mm $\times$ 100 mm of open area and a frame of 5 mm on all sides.  The operational solid fraction $s$ and the number of layers  $N$ are free parameters to be adjusted.  The ACE should be measured at a free stream velocity close to 5 m s$^{-1}$ and in the presence of fog.   }
 \label{S3}
\end{figure}

\newpage

\singlespacing
\noindent Video S1: Fog flow through and around a four-layer harp collector captured at a rate of 4000 frames per second.  The open area of the collector is 100 mm $\times$  100 mm and the wind velocity is approximately 4 m s$^{-1}$

\newpage

\doublespacing

%% MODELS FOR THE FILTERED FRACTION
\section{Models for the filtered fraction}
We consider below three alternative models for predicting the filtered fraction $\varphi$ for a fog collector constituted of $N$ layers, each with operational solidity $s$.  As stated in the main text, the approach taken by most models is based on the following relation for the filtered fraction:
 \begin{equation}
 \varphi= \frac{A_\infty}{A}=\frac{u}{u_\infty} = \sqrt{\frac{C_D}{k} } 
 \end{equation}
Therefore, we seek to express the drag coefficient $C_D$ and pressure drop coefficient $k$ in terms of $N$ and $s$.

\subsection{Glauert1932 Model}
Glauert and coworkers \cite{Glauert1932} presented one of the first detailed analysis of the flow through and around a porous structure. Treating the flow in the porous medium as a series of sources, they arrived at the following relations:
\begin{equation}
C_D =\frac{k}{(1 + \frac{1}{4}k)^2} 
\end{equation}
and  
\begin{equation}
 k = s \left( \frac{1}{(1-s)^2} - \frac{2}{3} \right)
 \end{equation} 
although the equation for $C_D$ never appears in this form in their paper.  The first relation was re-affirmed by Taylor \cite{Taylor1944a} using two different approaches.  However, as was clear at the time, the relation does not admit drag coefficients greater than 1 even in the limit of $k$ approaching infinity (a solid plate) while it is known that the drag coefficient for a square plate is in fact 1.18 in the range of $Re$ numbers of interest.  Luckily, the equation is most robust for small $k$ (small solid fraction), which is the regime of interest for fog collectors.  Taylor and Davies \cite{Taylor1944} state that the equation could be valid for $k \le 4$.

\subsection{Rivera2011 Model}
Rivera \cite{Rivera2011} took a slightly different approach by considering the flow through and around the collectors as the superposition of two distinct flow fields with the condition $u_\infty = u +\hat u$, where $u_\infty$ is the velocity of the unperturbed upstream flow, $u$  is the velocity of the uniform flow that filters through the porous collector and $\hat u$ is the velocity of the flow associated with a solid collector.  Rivera then equates the pressure drop for the two components of the flow field based on Bernoulli's principle:
\begin{equation}
\Delta p = \frac{\rho_{air} \hat u ^2}{2}\hat C_D =  \frac{\rho_{air} u^2}{2}k
\end{equation}
and since  $\hat u = u_\infty - u$, we have:
\begin{equation}
\frac{\rho_{air}(u_\infty - u) ^2}{2}\hat C_D =  \frac{\rho_{air} u^2}{2}k
\end{equation}
rearranging gives: 
\begin{equation}
\left(\frac{k}{\hat C_D} \right )^{1/2} =  \frac{u_\infty - u}{u}
\end{equation}
and finally,
\begin{equation}
\varphi =\frac{1}{1+(k/\hat C_D)^{1/2}}
\end{equation}
where $\hat C_D = 1.18$ is the drag coefficient corresponding to a solid ($s = 1$) collector with square aspect ratio.  For the pressure drop coefficient, the empirical relation given by Idel'Cik \cite{Idel'Cik1969}  was selected:
\begin{equation}
k = 1.3s + \left( \frac{s}{1-s} \right)^2
\end{equation}

\subsection{Koo1973 Model}
Koo and James \cite{Koo1973} revisited the model of Taylor and Davies \cite{Taylor1944} by considering the flow through a porous medium as equivalent to distributed sources.  The problem was solved so as to ensure conservation of mass and momentum across the mesh, leading to the implicit relations:
\begin{eqnarray}
k  &=& \frac{2Dk + (Dk)^2}{(1 + Dk)^2} \left (1 + \frac{Dk}{2} \right )^2 \\
C_D &=& \frac{k}{(1+\frac{1}{2}Dk)^2}
\end{eqnarray}
where $D$ is the source strength.  Because, Koo and James\cite{Koo1973} were mostly concerned about the relation between $k$ and $C_D$, they did not try to express $k$ in terms of the solidity.  We can however use  Idel'Cik's \cite{Idel'Cik1969} empirical relation (Eq.~8)  to close the problem.

\newpage

\bibpunct{[S}{]}{,}{n}{}{;}

%\bibliography{Collectors_supp} 

\end{document}